\documentstyle[psfig,11pt]{article}

\title{Approximate Quantum Fourier Transform and Decoherence
\\[1cm]
\normalsize
Adriano Barenco$^a$, Artur Ekert$^a$, Kalle-Antti Suominen$^b$
  and
P\"aivi T\"orm\"a$^c$ \\[0.5cm]
\small
$^a$Clarendon Laboratory, Department of Physics, University
  of Oxford, Parks Road, OX1 3PU Oxford, U.K.\\ 
$^b$Theoretical
  Physics Division, Department of Physics, University of Helsinki, PL
  9, 00014 Helsingin yliopisto, Finland\\ 
$^c$Research Institute for
  Theoretical Physics, University of Helsinki, PL 9, 00014 Helsingin
  yliopisto, Finland}

\begin{document}

\maketitle

\begin{abstract}
  We discuss the advantages of using the approximate quantum Fourier
  transform (AQFT) in algorithms which involve periodicity
  estimations. We analyse quantum networks performing AQFT in the
  presence of decoherence and show that extensive approximations can
  be made before the accuracy of AQFT (as compared with regular
  quantum Fourier transform) is compromised.  We show that for some
  computations an approximation may imply a better performance.
\\[1cm]
PACS: 89.70.+c, 03.65.-w, 42.50.Lc \hfill Submitted to Phys. Rev. A.
(Jan. 96)
\end{abstract}

\baselineskip=5mm

\section{Introduction}\label{introduction}

In the course of history many ingenious mechanical, acoustic and
optical devices have been invented for performing Fourier 
transforms~\cite{Fourier} 
(including nature's own such as the human ear). Most of them are now
of merely historical interest since the arrival of the computer--based
algorithm known as the fast Fourier transform
(FFT)~\cite{Cooley,Brigham} which efficiently computes the discrete
Fourier transform.  The FFT algorithm can also be phrased in terms of
quantum dynamics, i.e., in terms of unitary operations
performed by a quantum computer on quantum registers.  Indeed, all
known quantum algorithms employ the quantum version of Fourier
transforms, either explicitly or indirectly.  It is used for the
periodicity estimation in the Shor algorithm~\cite{Shor94} and its
approximate version (the Hadamard transform) is commonly used to
prepare quantum registers in coherent superpositions of different
values.

In this paper we analyse the performance of the quantum Fourier
transform (QFT) in the presence of decoherence. In particular we show
that as far as the periodicity estimation is concerned the approximate
quantum Fourier transform (AQFT) can yield better results than the
full Fourier transform.

In the following we use some terms which were originally adopted from
the classical theory of information and computer science and became
standard in the lore of quantum computation. More detailed
descriptions can be found e.g.~in
Refs.~\cite{BDEJ95,SW95,Vedral96} and in some recent reviews~\cite{BDL95}.
\begin{itemize}
\item A {\em qubit} is a two--state quantum system; it has a chosen
  `computational basis' $\{|0\rangle,|1\rangle\}$ corresponding to the
  classical bit values $0$ and $1$. Boolean operations which map
  sequences of 0's and 1's into another sequences of 0's and 1's are
  defined with respect to this computational basis. A collection of
  $L$ qubits is called a {\em register} of size $L$.
\item Information is stored in the registers in binary form. For
  example, number $6$ is represented by a register in state $|1\rangle
  \otimes |1\rangle \otimes |0\rangle$. In more compact notation:
  $|a\rangle$ stands for the direct product $|a_{L-1}\rangle \otimes
  |a_{L-2}\rangle\ldots|a_1\rangle \otimes|a_0\rangle$ which
  represents a quantum register prepared with the value $a=2^0 a_0+
  2^1 a_1 +\ldots 2^{L-1} a_{L-1}$.
\item A {\em quantum logic gate} is an elementary quantum computing
  device which performs a fixed unitary operation on selected qubits
  in a fixed period of time.
\item A {\em quantum network} is a quantum computing device consisting
  of quantum logic gates whose computational steps are synchronised in
  time. The outputs of some of the gates are connected by wires to the
  inputs of others.  The {\em size} of the network is its number of
  gates.
\item A {\em quantum computer} will be viewed here as a quantum
  network (or a family of quantum networks). Quantum computation is
  defined as a unitary evolution of the network which takes its
  initial state ``input'' into some final state ``output''.
\end{itemize}

Our presentation starts with a brief mathematical introduction to the
approximate discrete Fourier transform which is followed by the
description of its quantum implementation in terms of quantum
networks. Then we analyse how the performance of the QFT in the
periodicity estimation is affected by the approximations in the
algorithms and by decoherence. We also comment on possible
simplifications in practical implementations of quantum networks
effecting the QFT and AQFT. The quantum algorithm for the fast Fourier
transform which we use in this paper was originally proposed by
Coppersmith and by Deutsch (independently)~\cite{C94}.

\section{Discrete Fourier Transforms}\label{transforms}

The discrete Fourier transform is a unitary transformation of a
$s$--dimensional vector $\{f(0), f(1), f(2),\ldots,f(s-1)\}$ defined
by:
\begin{equation}
   \tilde f(c)=\frac{1}{\sqrt s}\sum_{a=0}^{s-1}e^{2\pi i ac/s} f(a),
   \label{def}
\end{equation}
where $f(a)$ and $\tilde f(c)$ are in general complex numbers. It can
also be represented as a unitary matrix
\begin{equation}
  \frac{1}{\sqrt{s}} \left(\begin{array}{ccccc} 1 & 1 & 1 & \ldots &
  1\\ 1 & \omega & \omega^2 & \ldots & \omega^{(s-1)} \\ 1 & \omega^2
  & \omega^4 & \ldots & \omega^{2(s-1)}\\ 1 & \omega^3 & \omega^6 &
  \ldots & \omega^{3(s-1)}\\ \vdots & \vdots & \vdots & \ddots &
  \vdots \\ 1 & \omega^{(s-1)} & \omega^{2(s-1)} & \ldots &
  \omega^{(s-1)^2}
   \end{array}\right),
\end{equation}
where $\omega=\exp(2\pi i /s)$ is the $s$th root of unity. In the
following we assume that $s$ is a power of $2$, i.e., $s=2^L$
for some $L$; this is a natural choice when binary coding is used. The
approximate discrete Fourier transform can be conveniently described
when we write the product $ac$ in the exponent on the r.h.s.\ of Eq.
(\ref{def}) in the binary notation.  Writing
\begin{equation}
  a=\sum_{i=0}^{L-1}a_i2^i \quad\quad ;\quad\quad
  c=\sum_{i=0}^{L-1}c_i2^i
\end{equation}
we obtain
\begin{eqnarray}
  ac &=& a_0c_0 + (a_0c_1+a_1c_0) 2 + (a_0c_2 + a_1c_1 + a_2c_0) 2^2 +
  \ldots\nonumber\\ &+& (a_0c_{L-1} +\ldots +a_{L-1}c_0) 2^{L-1} +
  {\cal O}(2^L).
\end{eqnarray}
Because $\omega^x=1$ for $x\ge s$, the terms ${\cal O}(2^L)$ do not
contribute to the transform, and the term $\exp(2\pi i ac/2^L)$ in
Eq.~(\ref{def}) can be expressed as
\begin{eqnarray}
   \exp(2\pi i ac/2^L) &=& \exp(2\pi i (a_0c_0)/2^L)\exp(2\pi i
      (a_0c_1+a_1c_0)/2^{L-1})\dots\nonumber\\ & &\times\exp(2\pi i 
      (a_0c_{L-1} +\ldots +a_{L-1}c_0)/2) .
\label{aprx}
\end{eqnarray}
Beginning from the right of this expression, the arguments in the
exponentials become smaller and smaller. In the approximate Fourier
transform parameterised by an integer $m$, the $L-m$ smallest terms
are neglected. In all the remaining terms the arguments are then
multiples of $2\pi/2^m$. The $2^m$th root of unity becomes the basic
element of the approximate Fourier transform as opposed to $2^L$th
root of unity which used in the ordinary Fourier transform.  (The
ordinary Fourier transform is obtained for $m=L$; when $m=1$ we obtain
the Hadamard transform, for which all terms but the last one are
dropped.)

The quantum version of the discrete Fourier transform is a unitary
transformation which can be written in a chosen computational basis
$\{ | 0 \rangle,| 1 \rangle,\ldots,| s-1 \rangle\}$ as
\begin{equation}
   \mbox{\rm QFT}_s : | a \rangle \longmapsto \frac{1}{\sqrt{s}} 
   \sum_{c=0}^{s-1}
   \exp(2\pi iac/s)\: | c \rangle  .
\label{qftdef}
\end{equation}
More generally, $\mbox{QFT}_s$ effects the discrete Fourier transform
of the input amplitudes. If
\begin{equation}
   \mbox{\rm QFT}_s: \sum_a f(a)| a \rangle \longmapsto
   \sum_c \tilde{f}(c)| c \rangle,
\end{equation}
then the coefficients $\tilde{f}(c)$ are the discrete Fourier
transforms of $f(a)$'s. This definition can be trivially extended to
cover the approximate quantum Fourier transform (AQFT).  We will
analyse the approximations involved in the AQFT in terms of
computational networks.

\section{Quantum Networks for AQFT}\label{networks}

Quantum networks for AQFT can be constructed following the description
of the fast Fourier transform algorithm (as described by
Knuth~\cite{Knuth}). This efficient classical algorithm needs to be
re--expressed in terms of unitary operations~\cite{C94}. The
construction requires only two basic unitary operations. The first
operation is a one--bit transformation $A_i$ (one--bit gate) that acts
on a qubit $q_i$ of the register and effects
\begin{equation}
  \begin{array}{l}
    | 0 \rangle\longrightarrow \frac{1}{\sqrt{2}} (| 0 \rangle+| 1
    \rangle) \\
    | 1 \rangle\longrightarrow \frac{1}{\sqrt{2}} (| 0 \rangle-| 1
    \rangle).
  \end{array}
  \mbox{\hspace{3cm}} 
  \mbox{
  \setlength{\unitlength}{0.030in}
  \begin{picture}(30,0)(0,15)
    \put(0,15){$q_i$} \put(5,15){\line(1,0){5}}
    \put(20,15){\line(1,0){5}} \put(10,10){\framebox(10,10){$A_i$}}
  \end{picture}
  }
\end{equation}
The diagram on the right provides a schematic representation of the
gate acting on a qubit $q$.  The second operation is a two--bit gate
$B_{jk}$ that effects
\begin{equation}
  \begin{array}{l}
    | 00 \rangle \longrightarrow | 00 \rangle  \\
    | 01 \rangle \longrightarrow | 01 \rangle  \\
    | 10 \rangle \longrightarrow | 10 \rangle  \\
    | 11 \rangle \longrightarrow \exp(i\theta_{jk})| 11 \rangle
   \end{array}
   \mbox{\hspace{1.5cm}} \mbox{
\setlength{\unitlength}{0.030in}
\begin{picture}(25,0)(0,20)
  \put(-2,15){$q_k$} \put(-2,30){$q_j$} \put(5,15){\line(1,0){5}}
  \put(20,15){\line(1,0){5}} \put(5,30){\line(1,0){20}}
  \put(15,30){\circle*{3}} \put(15,20){\line(0,1){10}}
  \put(10,10){\framebox(10,10){$\theta_{jk}$}}
\end{picture}
}
\equiv
\mbox{\hspace{0.2cm}} 
\mbox{
\setlength{\unitlength}{0.030in}
\begin{picture}(30,0)(0,20)
  \put(-2,15){$q_k$} \put(-2,30){$q_j$}
  \put(5,30){\line(1,0){5}}
  \put(20,30){\line(1,0){5}}
  \put(5,15){\line(1,0){20}}
  \put(15,15){\circle*{3}} \put(15,15){\line(0,1){10}}
  \put(10,25){\framebox(10,10){$\theta_{jk}$}}
\end{picture}
}
\end{equation}
where $\theta_{jk}$ depends on the qubits $j$ and $k$ on which the
gate acts and equals $\theta_{jk}=\pi/2^{k-j}$.  The transformation
$B_{jk}$ is an elementary two qubit operation which affects only
states with a $1$ in both position $j$ and $k$ regardless the state of
the remaining qubits. 

The QFT on a register of size $1$ reduces to a single operation $A$
performed on a single qubit (cf. equation (\ref{qftdef}) for $s=2$).
The extension of the QFT network to a register of any size $L$ follows
a simple pattern of gates $A$ and $B$ which can be seen in
Fig.~\ref{DFT}. It shows the QFT network operating on four qubits
which can be written as the sequence of the following $10$ elementary
operations (read from left to right)
\begin{equation}
  (A_3)(B_{23}A_2)(B_{13}B_{12}A_1)(B_{03}B_{02}B_{01}A_0).
\end{equation}
The bit values at the output should be read in the reversed order 
(see Fig.~\ref{DFT}).

The number of gates needed to complete the QFT grows only as a
quadratic function of the size of the register: for a transformation
on a $L$ qubit register, we require $L$ operations $A$ and $L(L-1)/2$
operations $B$, in total $L(L+1)/2$ elementary operations. Thus the
quantum QFT can be performed efficiently.

The AQFT of degree $m$ is represented by a similar network in which
the two--bit gates that act on qubits which are far apart (in the
register) are neglected, i.e., those operations $B_{jk}$ for which the
phase shift $\theta_{jk}\equiv\pi/2^{k-j}<\pi/2^m$ for some $m$ such
that $1\le m\le L$ are dropped (cf. Eq.(\ref{aprx})). In that case, we
need $L$ operations $A$, and $(2L-m)(m-1)/2$ operations $B$, which is
an improvement on the QFT case since $m<L$. In Fig.~\ref{AQFT} we show
the $m=2$ AQFT network counterpart to the QFT network shown in
Fig.~\ref{DFT}. The matrix elements of the QFT and the AQFT differ by
a multiplicative factors of the form $\exp (i\epsilon)$ with
$|\epsilon|\le 2\pi L/2^m$. The execution time of the AQFT grows as
$\sim Lm$.

\section{Estimating Periodicity}\label{estimating}

The quantum Fourier transform, like the ordinary Fourier transform, is
a powerful tool for uncovering periodicities. Any periodicity in
probability amplitudes describing a quantum state of a register in a
computational basis can be estimated (with some probability of
success) by performing the QFT followed by a measurement of the
register in the computational basis. The result is obtained by reading
the qubits of the register in the reversed order.

For example, an interesting periodic state which plays an important role in
Shor's quantum factoring algorithm can be written as
\begin{equation}
  | \Psi \rangle=\frac{1}{\sqrt{{\cal N}}}\sum_{a=0}^{2^L-1} f(a) | a
  \rangle,
\label{init}
\end{equation}
where ${\cal N}$ is an appropriate normalisation factor and
\begin{equation}
  f(a)=\delta_{l,\;a \;\bmod \;r}.
  \label{periodic}
\end{equation}
It is the state of a quantum register of size $L$ in which the
probability amplitudes $f(a)$ occur with periodicity $r$ and offset
$l$.  If this offset is unknown, a measurement performed in the
computational basis cannot reveal $r$ or any of its integer multiples
directly. This is illustrated in Fig.~\ref{fourier}(a). However, if we
perform the QFT on the register first and subsequently measure its
state we obtain number $\bar c$ which, with probability greater than
$4/\pi^2$, is a multiple of $2^L/r$ regardless the offset $l$
(cf.~Appendix).  The probability is not equal to unity, because the
finite size of the register leads to a ``broadening" of the
Fourier--transformed data, as illustrated in Fig.~\ref{fourier}(b).
(This is because $2^L/r$ is not necessarily an integer, and the
quantum register can have only integer values; this is discussed in
detail in the Appendix).

For the AQFT the corresponding probability, in the limit of large $L$, 
satisfies 
\begin{equation}
   \mbox{Prob}_A\ge \frac{8}{\pi^2} \sin^2 \left( \frac{\pi}{4} \frac{m}{L}
     \right).
\end{equation}
This result is derived in the Appendix. The effect of the
approximation is illustrated in Fig.~\ref{approximate} where we plot
the modulus of the amplitude of the transformed state $| \Psi \rangle$
(with $l=9$ and $r=10$) for the AQFT of different orders $m$.
Fig.~\ref{phases} shows how the phase of the transformed state becomes
corrupted when $m$ becomes smaller.

If the quantum Fourier transform forms a part of a randomised
algorithm then the computation can be repeated several times in order
to amplify the probability of the correct result. In such cases the
performance of the AQFT is only polynomially less efficient than that
of the QFT.  For example, consider Shor's quantum factoring algorithm
and substitute the AQFT for the QFT. In order to obtain a correct
factor with a prescribed probability of success, we have to repeat the
computation several times. Let $k$ and $k'$ be the number of runs
respectively with the QFT and the AQFT so that we obtain the same
probability of getting at least one correct result, i.e.,
\begin{equation}
   1-(1-p)^k = 1-(1-p')^{k'} .
\end{equation}
Here $p=4/\pi^2$ and $p'=\frac{8}{\pi^2} \sin^2
\left(\frac{4}{\pi}\frac{m}{L}\right)$ are the corresponding
probabilities of success in a single run. The ratio $k'/k$ scales as
\begin{equation}
  \frac{k'}{k} = \frac{\log \left(1-\frac{4}{\pi^2}\right)}{\log
    \left(1-\frac{8}{\pi^2} \sin^2 \left(
    \frac{4}{\pi}\frac{m}{L}\right) \right)} < C
  \left(\frac{L}{m}\right)^3
\end{equation}
for some $C$ (the upper bound is found graphically). 
This shows that in the quantum factoring algorithm the AQFT is not
less efficient than the ordinary QFT, i.e., the ratio $k'/k$ scales only 
polynomially with $L/m$. Moreover, we will show that in
the presence of decoherence the AQFT can perform better than the QFT
even in a single computational run!

\section{Decoherence}\label{decoherence}

Quantum computation requires a controlled, quantum--mechanically
coherent evolution at the level of individual quantum systems such as
atoms or photons. This imposes severe requirements on quantum computer
hardware. If we are to harness the unique power of quantum computers,
such systems will have to be manufactured with unprecedented
tolerances and shielded from noise to an unprecedented degree. Even a
minute interaction with the environment will lead to a non--unitary
evolution of the computer and its state will, in general, evolve into
a mixed state described by a reduced density operator $\rho$, which is
obtained from the density operator $\rho_{\rm total}$ of the total
computer+environment system by taking a trace over all the quantum
states of the environment:
\begin{equation}
  \rho = {\rm Tr}_{\rm environment}\, (\rho_{\rm total}).
\end{equation}
Consider, for example, a quantum register of size $L$ which is
prepared initially in some pure state and then left on its own. As
time goes by, the qubits become entangled with the environment. Both
the diagonal and the off--diagonal elements of the density matrix
$\rho$ (expressed in a computational basis) are usually affected by
this process (cf.~\cite{Unruh95,Palma96}). The rate of change of the
diagonal and the off--diagonal elements depends on the type of
coupling to the environment, however, there are realistic cases where
the dissapearance of the off--diagonal elements, known as decoherence,
takes place on much faster time scale. In this case a simple
mathematical model of decoherence has been proposed~\cite{Zurek91}. It
assumes that the environment effectively acts as a measuring
apparatus, i.e., a single qubit in state $c_0 | 0 \rangle +c_1 | 1
\rangle$ evolves together with the environment as
\begin{equation}
  (c_0 | 0 \rangle + c_1| 1 \rangle) |a \rangle \longrightarrow
    c_0| 0 \rangle |a_0 \rangle + c_1 | 1 \rangle |a_1\rangle,
\end{equation}
where states $|a\rangle, |a_0\rangle, |a_1\rangle$ are the states of the
environment and $|a_0\rangle, |a_1\rangle$ are usually not orthogonal
($\langle a_0| a_1\rangle \ne 0$).  The elements of the density matrix
evolve as
\begin{equation}
\rho_{ij}(0)=c_i(0)c^*_j(0) \longrightarrow \rho_{ij}(t) =c_i(t)c^*_j(t)
\langle a_i(t)| a_j(t)\rangle , \qquad i,j =0,1.
\end{equation}
States $|a_0\rangle$ and $|a_1 \rangle$ become more and more
orthogonal to each other whilst the coefficients $\{c_i\}$ remain
unchanged.  Consequently the off--diagonal elements of $\rho$
disappear due to the $\langle a_0(t) | a_1(t)\rangle$ factor and
the diagonal elements are not affected.

There is an alternative way of thinking about this process. The
environment is regarded as a bosonic heat bath, which introduces phase
fluctuations to the qubit states, i.e., it induces random phase
fluctuations in the coefficients $c_0$ and $c_1$ such that
\begin{equation}
   c_0\left|\, 0\right\rangle + c_1\left|\, 1\right\rangle\rightarrow
   c_0e^{-i\phi}\left|\, 0\right\rangle + c_1e^{i\phi}\left|\, 1\right\rangle.
\end{equation}
The direction and the magnitude of each phase fluctuation $\phi$ is
chosen randomly following the Gaussian distribution
\begin{equation}
  P(\phi)d\phi = \frac{1}{\sqrt{2\pi}\delta} \exp\left[-\frac{1}{2}
  \left(\frac{\phi}{\delta}\right)^2 \right]d\phi,
\end{equation}
where the distribution width $\delta$ defines the strength of the
coupling to the quantum states of the environment. The elements of the
density matrix $\rho$ are then reconstructed as $\rho_{ij} = \langle
c_i c_j^* \rangle $, where the average $\langle \rangle$ is taken over
different realisations of the phase fluctuations within a given period
of time (cf.~\cite{Palma96}). The diagonal elements do not
depend on $\phi$, whereas the off--diagonal term $\langle
c_0c_1e^{i2\phi} \rangle$ averages to zero for a sufficiently long
period of time. 

The latter approach to decoherence is very convenient for numerical
simulations and was chosen for the purpose of this paper. It is
similar to the Monte Carlo wave function method used in quantum
optics~\cite{MCWF}.

\section{AQFT and Decoherence}\label{aqft+deco}

We have analysed decoherence in the AQFT networks assuming that the
environment introduces a random phase fluctuation in a qubit
probability amplitudes each time the qubit is affected by gate $B$. In
our model we have not attached any decoherence effects to gate $A$. In
most of the suggested physical realisations the single qubit
operations are quite fast, whereas the conditional logic needed in
two--qubit operations is often much harder to produce, which makes
these operations slower than the single qubit ones, and often much
more susceptible to decoherence.  For instance, in the ion trap model
proposed by Cirac and Zoller~\cite{Cirac95} single qubit operations
require only one laser pulse interacting with one atom, whereas in a
two--qubit operation two subsequent laser pulses are needed, and the
atoms involved must form an entangled state with a trap phonon mode
between the pulses. In any case, introducing decoherence to gate $A$
operations and `wires' in the network would not affect our results
much, only the time scale for decoherence would change.
 
We have quantified the performance of the AQFT by introducing the
quality factor $Q$. It is simply the probability of obtaining an
integer which is closest to any integer multiple of $2^L/r$, when
the state of the register is measured after the transformation.  In
the decoherence free environment analysed in Sec.~\ref{estimating} we
obtain $Q=1$ for integer values of $2^L/r$ and for a randomly selected
$r$ the quality factor $Q$ is of the order $4/\pi^2$ for the QFT and
of the order of $\frac{8}{\pi^2} \sin^2
\left(\frac{4}{\pi}\frac{m}{L}\right)$ for the AQFT of degree $m$.

In Fig.~\ref{decoh1} we show how the quality factor $Q$ behaves as a
function of $m$ and $\delta$ (which characterises the strength of the
coupling to the environment). For $\delta >0$ the maximum of $Q$ is
obtained for $m<L$. Thus in the presence of decoherence one should use
the AQFT rather than the QFT.

This `less is more' result can be easily understood. The AQFT means
less gates in the network and because each $B$ gate introduces phase
fluctuations the approximate network generates less decoherence as
compared to the regular QFT network. By decreasing $m$ we effectively
decrease the impact of decoherence. On the other hand decreasing $m$
implies approximations which reduces the quality factor. This
trade-off between the two phenomena results in the maximum value $Q$
for $m\in [1,L]$.

It is worth pointing out that for $\delta=0$ (no decoherence) $Q$ remains
almost constant for those values of $m$ that satisfy the lower bound
condition (derived in the Appendix)
\begin{equation}
   m > \log_2 L + 2  .  \label{lowerB}
\end{equation}
and when $\delta >0$ the optimum $m$ is found
near this lower bound. In Fig.~\ref{decoh2} we also show how $Q$
decreases rapidly with $L$ in the QFT network (although there is not
enough data in the figure to determine if it really decreases
exponentially).

Our simulations were performed for ensembles which consisted typically
of a one to two thousands individual realisations.

\section{Conclusions and comments}

We have analysed the approximate quantum Fourier transform in the
presence of decoherence and found that the approximation does not to
imply a worse performance. On the contrary, using the periodicity
estimation as an example of a computational task, we have shown that
for some algorithms the approximation may actually imply a better
performance.

Needless to say, there is room for further simplifications of the
quantum Fourier transform which may lead to at least partial
suppressing of unwelcome effects of decoherence. For example, if the
QFT is followed by a bit by bit measurement of the register then the
conditional dynamics in the network can be converted to a sequence of
conditional bit by bit measurements (cf.~\cite{CMU}). Here, we wanted
to show that there are cases where quantum networks that are composed
of imprecise components can guarantee a `pretty good performance'.
This topic has also a more general context; it has been shown that
reliable classical networks can be assembled from unreliable
components~\cite{Feige94}. It is an open question whether a similar
result holds for quantum networks.

\subsection*{Acknowledgements}

A. B. thanks E. S. Trounz for discussions.  A. E. is supported by the
Royal Society. A. B. acknowledges the Berrow fund at Lincoln College
(Oxford), and K.-A. S. and P. T. acknowledge the Academy of Finland
for financial support, and thank A. E. and the rest of the group for
kind hospitality during visits to Oxford.

\appendix

\section*{}

Consider the quantum state
\begin{equation}
  | \Psi \rangle=\frac{1}{\sqrt{{\cal N}}}\sum_{a=0}^{2^L-1} f(a) | a
  \rangle,
  \label{init2}
\end{equation}
where 
\begin{equation}
  f(a)=\delta_{l,\;a \;\bmod \;r}. \label{periodic2}
\end{equation}
Here $f(a)$ is a periodic function with period $r \ll 2^L$ and offset
$l$, which is an arbitrary positive integer smaller than $r$; see
Fig.~\ref{fourier}(a).  The normalization factor is equal to the
number of non--zero values $f(a)$: ${\cal N}=[2^L/r]$. Because $r \ll
2^L$ we use from now on $[2^L/r]\simeq 2^L/r$.  The state
(\ref{init2}) plays an important role in the Shor quantum
factorisation algorithm (the algorithm enables us to factorise an
integer $N$ by finding $r$ such that $x^r=(1 \bmod N)$ for some $x$
coprime with $N$ --- $r$ is estimated from a quantum computation that
generates a state of the form~(\ref{init2}).)
 
Applying QFT to this state we obtain
\begin{equation}
   | \tilde\Psi\rangle= \sum \tilde f(c)| c \rangle,
\end{equation}
where
\begin{equation}
   \tilde{f}(c) = \frac{\sqrt{r}}{2^L} \sum_{j=0}^{2^L/r-1} 
   \exp(2\pi i(jr+l)c/2^L ).
\end{equation} 
The probability of seeing an integer $c$ is then
\begin{equation}
  \mbox{Prob}(c) = |\tilde f(c)|^2=\frac{r}{2^{2L}} \left|
  \sum_{j=0}^{2^L/r-1} \exp(2\pi ij (rc \bmod 2^L)/2^L) \right| ^2.
  \label{prob_c}
\end{equation}
As it can be seen from Fig.~\ref{fourier}(b) the peaks of the power
spectrum of $f(a)$ are centered at integers $c$ which are the closest
approximation to multiplies of $2^L/r$.

Let us now evaluate $\mbox{Prob}(\bar{c})$ for $\bar{c}$ being the
closest integer to $\lambda 2^L/r$, i.e., $\bar c=[ \lambda
2^L/r]$. By definition $\bar{c}$ must satisfy
\begin{equation}
  -\frac{1}{2} < \bar{c}-\lambda \frac{2^L}{r} < \frac{1}{2} .
\label{condition}
\end{equation}
We define $\theta_{\bar{c}}=2 \pi (r \bar{c} \bmod \;2^L)$ so that
$\mbox{Prob}(\bar{c})$ now involves a geometric series with ratio
$\exp (i\theta_{\bar{c}})$. By viewing these terms as vectors on an
Argand diagram it is clear that the total distance from the origin
decreases as $\theta_{\bar{c}}$ increases. Hence $\mbox{Prob}(\bar{c})
\geq \mbox{Prob}(\mbox{ $\bar{c}$ with largest allowed }
\theta_{\bar{c}})$.  Let us denote by $\theta_{\max}$ the largest
allowed $\theta_{\bar{c}}$. By Eq.~(\ref{condition}), $\theta_{\max}
\leq \pi r/2^L$ and summing the geometric series with $\theta_{\max} =
\pi r/2^L$ (see Fig.~\ref{Argand1}) we obtain
\begin{equation}
  \mbox{Prob}(\bar{c}) \geq \frac{r}{2^{2L} } \frac{1}{\sin^2 \left(
  \frac{\pi}{2} \frac{r}{2^L}\right)}\simeq\frac{4}{\pi^2 } \frac{1}{r},
  \label{cprime}
\end{equation}
where we have used the fact that $r/2^L$ is small. Since there are $r$ such
values $\bar{c}$, the total probability of seeing one of them is 
\begin{equation}
   \mbox{Prob} \geq 4/\pi^2  .
\end{equation}
By performing this measurement several times on different states $|
\Psi \rangle$ (each one with possibly different $l$), one gets with
high probability values $\bar{c}_0$, $\bar{c}_1, \ldots$ that are the
closest integers to $n_0 2^L/r$, $n_1 2^L/r, \ldots$ and which allow
to calculate $r$ [cf. inset in Fig.~\ref{fourier}(b)].

We estimate now the probability of measuring one of the desired values
$\bar{c}$ when the AQFT of order $m$ has been performed instead of the
QFT.  The difference between the QFT and the AQFT of order $m$ is in
the arguments of the exponentials in Eq. (\ref{prob_c}). The phase
difference for each term in the sum is
\begin{equation}
  \Delta(a,c)=\frac{2 \pi}{2^L} \left(a c 
  - \sum_{(j,k)\in {\cal E}}^{L-1} a_j c_k 2^{j+k} \right)  ,
\end{equation}
where
\begin{equation}
  {\cal E}=\left\{(j,k) \;|\; 0 \leq j,k \leq L-1,\; L-m\leq j+k \leq
  L-1 \right\}.
\end{equation}

The probability to measure $| \bar{c}\rangle$, where $\bar{c}$ is the
closest integer to one of the $r$ values $n 2^L/r$ now becomes
\begin{equation}
  \mbox{Prob}_A(c)= \frac{r}{2^{2L}} \left| \sum_{j=0}^{2^L/r-1}
  \exp(2\pi ij (rc \bmod 2^L)/2^L-i\Delta(jr,c)) \right| ^2.
\end{equation}
This is the same summation as is involved in the QFT, except that in
the case of the AQFT, each vector of the Argand diagram of
Fig.~\ref{Argand1}(a) may be rotated by an angle $\Delta(jr,c)$, as
shown in Fig.~\ref{Argand2}. In the worst case, when $a=c=2^L-1$, 
i.e., $a_i=c_i=1$ $\forall i$, $\Delta(a,c)$ is equal to
\begin{equation}
   \Delta_{\max}=\frac{2 \pi}{2^m}(L-m-1+2^{m-L}) .
\end{equation}
However, for any other values of $a$ and $c$, $0\leq\Delta(a,c) <
\Delta_{\max}$.  

We are interested in the lower bound for the probability so we assume
that the vectors in the Argand diagram fill one half of the circle
($\theta_{\max} = \pi r/2^L$) as illustrated in Fig. \ref{Argand1}(b).
The approximation allows to rotate each vector by the maximum angle
$\Delta_{\max}$. The minimum of the probability is obtained when half
of the vectors are rotated by $\Delta_{\max}$, see Fig.
\ref{Argand2}(b).  In this case vectors in two areas of size
$\Delta_{\max}$ cancel each other, and all we have to do is to
calculate geometrical sums of the vectors in the two areas of size
$\pi /2 - \Delta_{\max}$. In an area of that size there are
$\frac{2^L}{r} \left(\frac{1}{2}-\frac{\Delta_{\max}}{\pi}\right)$
vectors, since the total number of vectors is $2^L/r$. Note that
because $2^L\gg r$, we can assume that $[2^L/r]\pm 1 \simeq 2^L/r$.
The square of the geometric sum then becomes
\begin{eqnarray}
  \left| \sum_{j=0}^{\frac{2^L}{r} \left(
    \frac{1}{2}-\frac{\Delta_{\max}}{\pi} \right) - 1} \exp \left( i
  \frac{\pi r}{2^L} j\right) \right|^2 = \frac{\sin^2 \left( \frac{1}{2}
    \left( \frac{\pi}{2} - \Delta_{\max} \right) \right)} {\sin^2
    \left( \frac{\pi}{2}\frac{r}{2^L}\right)} .  \label{eqq1}
\end{eqnarray}

The two sum vectors in the two areas of size $\pi /2 - \Delta_{\max}$
are of equal length and orthogonal to each other, so the square of
their sum vector, contributing to the total probability, is twice the
value given by (\ref{eqq1}). Finally we obtain
\begin{equation}
  \mbox{Prob}_A \ge 2 \frac{r^2}{2^{2L}} \frac{\sin^2 \left(
    \frac{1}{2} \left( \frac{\pi}{2} - \Delta_{\max} \right) \right)}
  {\sin^2 \left( \frac{\pi}{2}\frac{r}{2^L}\right)} \simeq
  \frac{8}{\pi^2} \sin^2 \left( \frac{1}{2} \left(\frac{\pi}{2} -
  \Delta_{\max} \right)\right).
\label{probapp}
\end{equation}
For $\Delta_{\max} = 0$ this expression reduces to the result derived
for the QFT:
\begin{equation}
   \mbox{Prob}_A \ge \frac{8}{\pi^2} \sin^2 \left( \frac{\pi}{4} \right) =
   \frac{4}{\pi^2}  ,
\end{equation}
and for $\Delta_{\max} = \pi / 2$ we have $\mbox{Prob}_A \ge 0$.
To avoid a zero probability, $\Delta_{\max}$ must always be bounded by
\begin{equation}
   \Delta_{\max} = \frac{2\pi}{2^m} (L-m-1+2^{m-L}) < \frac{\pi}{2} ,
\end{equation}
which for large $L$ implies
\begin{equation}
   m > \log_2 L + 2  . 
\end{equation}
Eq. (\ref{lowerB}) gives a lower bound to the order of the AQFT
performed on a register of length $L$, if one wants to have a
non--zero probability of success in measuring a value $\bar{c}$.
Simple geometric considerations also show that $\Delta_{\max} < \pi
/2$ is a limit for a non--negligible probability: for $\Delta_{\max} >
\pi /2$ the vectors in the Argand diagram can be rotated so that there
is no longer any constructive interference, see Fig. \ref{Argand3}.

For large $L$ we can write
\begin{equation}
   \Delta_{\max} \simeq \frac{2\pi}{2^m} (L-m). \label{eqq2}
\end{equation}
If we use the lower bound for $m$ (\ref{lowerB}), we obtain
\begin{equation}
  \Delta_{\max} \le \frac{\pi}{2} \left( 1-\frac{m}{L}\right),
  \label{eqq3}
\end{equation}
which allows to write the probability~(\ref{probapp}) in a simple form
\begin{equation}
  \mbox{Prob}_A \ge \frac{8}{\pi^2} \sin^2 \left( \frac{\pi}{4}
  \frac{m}{L} \right).\label{ProbA}
\end{equation}


\setlength{\unitlength}{0.030in}
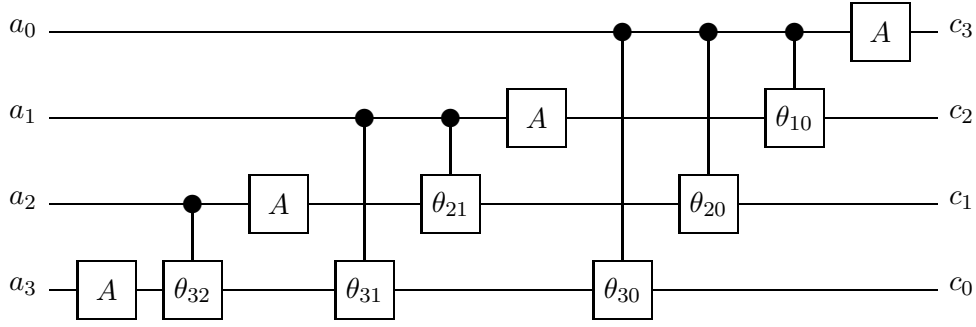
\begin{figure}
\begin{center}
\begin{picture}(150,65)
  \put(-2,15){$a_3$} \put(-2,30){$a_2$} \put(-2,45){$a_1$}
  \put(-2,60){$a_0$} \put(10,10){\framebox(10,10){$A$}}
  \put(25,10){\framebox(10,10){$\theta_{32}$}}
  \put(40,25){\framebox(10,10){$A$}}
  \put(55,10){\framebox(10,10){$\theta_{31}$}}
  \put(70,25){\framebox(10,10){$\theta_{21}$}}
  \put(85,40){\framebox(10,10){$A$}}
  \put(100,10){\framebox(10,10){$\theta_{30}$}}
  \put(115,25){\framebox(10,10){$\theta_{20}$}}
  \put(130,40){\framebox(10,10){$\theta_{10}$}}
  \put(145,55){\framebox(10,10){$A$}} \put(5,15){\line(1,0){5}}
  \put(20,15){\line(1,0){5}} \put(35,15){\line(1,0){20}}
  \put(65,15){\line(1,0){35}} \put(110,15){\line(1,0){50}}
  \put(5,30){\line(1,0){35}} \put(50,30){\line(1,0){20}}
  \put(80,30){\line(1,0){35}} \put(125,30){\line(1,0){35}}
  \put(5,45){\line(1,0){80}} \put(95,45){\line(1,0){35}}
  \put(140,45){\line(1,0){20}} \put(5,60){\line(1,0){140}}
  \put(155,60){\line(1,0){5}} \put(30,30){\circle*{3}}
  \put(30,20){\line(0,1){10}} \put(60,45){\circle*{3}}
  \put(60,20){\line(0,1){25}} \put(75,45){\circle*{3}}
  \put(75,35){\line(0,1){10}} \put(105,60){\circle*{3}}
  \put(105,20){\line(0,1){40}} \put(120,60){\circle*{3}}
  \put(120,35){\line(0,1){25}} \put(135,60){\circle*{3}}
  \put(135,50){\line(0,1){10}}
\put(162,60){$c_3$}
\put(162,45){$c_2$}
\put(162,30){$c_1$}
\put(162,15){$c_0$}
\end{picture}
\end{center}
\caption[fo1]{QFT network operating on a four--bit register. The phases
  $\theta_{jk}$ that appear in the operations $B_{jk}$ are related to
  the ``distance'' of the qubits ($j-k$) and are given by
  $\theta_{jk}=\pi/2^{j-k}$. The network should be read form the left
  to the right: first the gate $A$ is effected on the qubit $a_3$,
  then $B(\phi_{32})$ on $a_2$ and $a_3$, and so on.}
\label{DFT}
\end{figure}

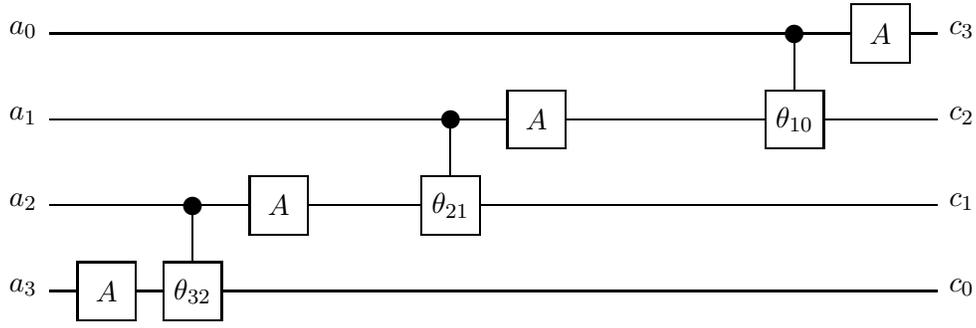
\begin{figure}
\begin{center}
\begin{picture}(150,65)
  \put(-2,15){$a_3$} \put(-2,30){$a_2$} \put(-2,45){$a_1$}
  \put(-2,60){$a_0$} \put(10,10){\framebox(10,10){$A$}}
  \put(25,10){\framebox(10,10){$\theta_{32}$}}
  \put(40,25){\framebox(10,10){$A$}}
  \put(70,25){\framebox(10,10){$\theta_{21}$}}
  \put(85,40){\framebox(10,10){$A$}}
  \put(130,40){\framebox(10,10){$\theta_{10}$}}
  \put(145,55){\framebox(10,10){$A$}} \put(5,15){\line(1,0){5}}
  \put(20,15){\line(1,0){5}} \put(35,15){\line(1,0){20}}
  \put(55,15){\line(1,0){45}} 
  \put(100,15){\line(1,0){60}}
  \put(5,30){\line(1,0){35}} \put(50,30){\line(1,0){20}}
  \put(80,30){\line(1,0){45}} 
  \put(115,30){\line(1,0){45}}
  \put(5,45){\line(1,0){80}} \put(95,45){\line(1,0){35}}
  \put(140,45){\line(1,0){20}} \put(5,60){\line(1,0){140}}
  \put(155,60){\line(1,0){5}} \put(30,30){\circle*{3}}
  \put(30,20){\line(0,1){10}} 
  \put(75,45){\circle*{3}}
  \put(75,35){\line(0,1){10}} 
  \put(135,60){\circle*{3}}
  \put(135,50){\line(0,1){10}}
\put(162,60){$c_3$}
\put(162,45){$c_2$}
\put(162,30){$c_1$}
\put(162,15){$c_0$}
\end{picture}
\end{center}
\caption[fo2]{
  AQFT network operating on a four--bit register when $m=2$.}
\label{AQFT}
\end{figure}

\begin{figure}
\caption[fo3]{
  (a) Function $f(a)$ in Eq.~(\ref{periodic}). The parameters are
  $l=8$, $r=10$ and the number of bits in the calculation is $L=9$ so
  that numbers up to $2^L-1=511$ can be encoded. (b) $|\tilde{f}(c)|$,
  obtained from $f(a)$ by a QFT. The inset shows $\bar{c}_3=155$ which
  is the closest integer to $3\cdot 512/10=153.6$. (See the Appendix
  for more details).}
\label{fourier}
\end{figure}

\begin{figure}
\caption[fo4]{
  Different orders of approximation in the AQFT performed on a state
  $| \Psi \rangle$ for which $f(a)=\delta_{\;9,\;a\; \bmod \;10}$. }
\label{approximate}
\end{figure}

\begin{figure}
\caption[fo5]{
  As Fig.~\protect\ref{approximate}, but showing the phase of the
  amplitudes, i.e., arg$(\tilde f(c))$.
\label{phases}}
\end{figure}

\begin{figure}
\caption[fo6]{The quality factor $Q$ as a function of $m$ for selected
  values of $\delta$. The register sizes are (a) $L=9$ and (b) $L=16$.
  Statistical errors are too small to be represented on the graph.
\label{decoh1}}
\end{figure}

\begin{figure}
\caption[fo7]{The quality factor $Q$ as a function of the register size $L$
  for QFT, with varying levels of decoherence, from $\delta=0.1$ (top
  line) to $\delta=0.5$ (bottom line). Statistical errors are too
  small to be represented on the graph.
\label{decoh2}
}
\end{figure}

\begin{figure}
\caption[ar1]{
  (a) Argand diagram corresponding to the sum of the phases that
  appear in the expression of $f(\bar{c})$ for $\bar{c}$ close to one
  of the values $n 2^L/r$. $\mbox{Prob}(\bar{c})$ is the norm of the
  vector resulting from the sum of each vector in the diagram. (b)
  $\mbox{Prob}(\bar{c})$ is bounded by the worst case situation in
  which we have taken $\theta_{\max}$ instead of $\theta_{\bar{c}}$,
  in this case the phases lie on an interval $[0, \pi]$ on the
  Argand diagram and a closed form expression can be found.}
\label{Argand1}
\end{figure} 

\begin{figure}
\caption[ar2]{
  (a) Argand diagram for the AQFT. Vectors are rotated by an angle
  $\Delta(jr,c)$.  (b) To obtain a closed form for a bound for
  $\mbox{Prob}_A(\bar{c})$ we consider the worst case in which half of
  the phases pick up a factor $\Delta_{\max}$.  }
\label{Argand2}
\end{figure}

\begin{figure} 
\caption[ar3]{
  In the case that the order $m$ of the AQFT is such that
  $\Delta_{\max}> \pi/2$, the individual phases can get scrambled in
  such a way that there is no constructive interference effect.  The
  probability $\mbox{Prob}_A(\bar{c})$ can become vanishingly small
  and the AQFT of order $m$ is inefficient.}
\label{Argand3} 
\end{figure} 

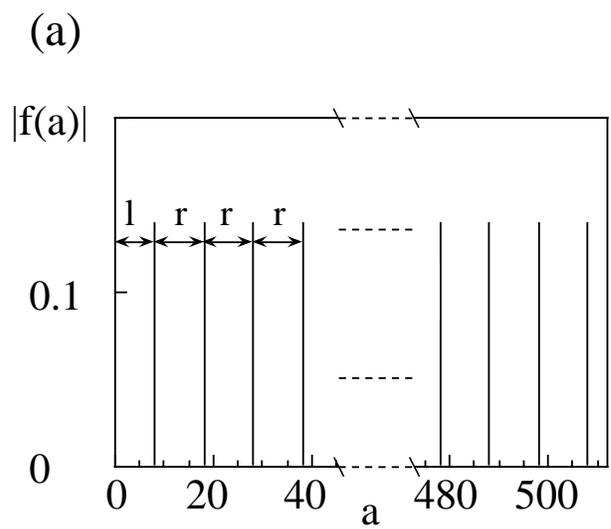
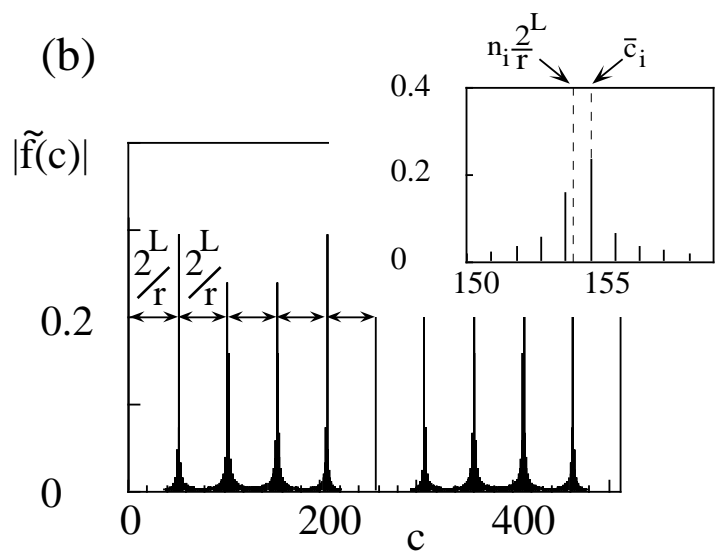
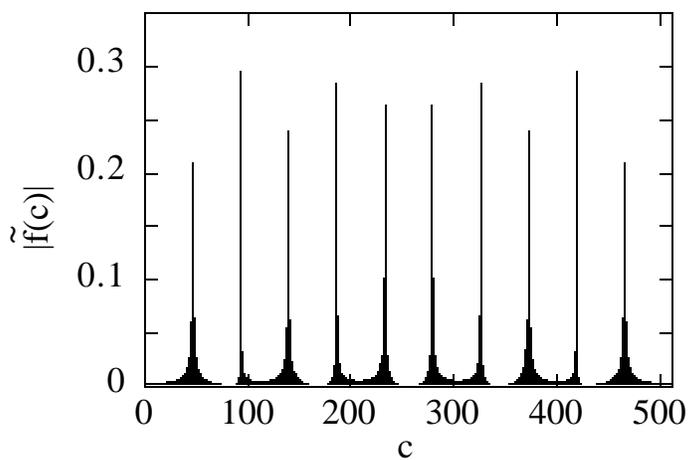
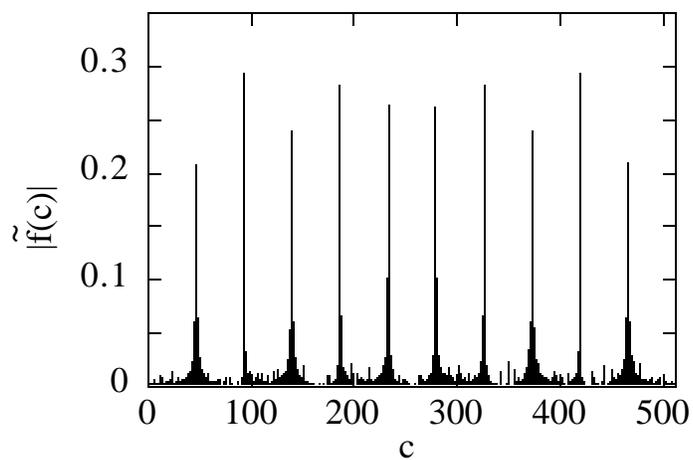
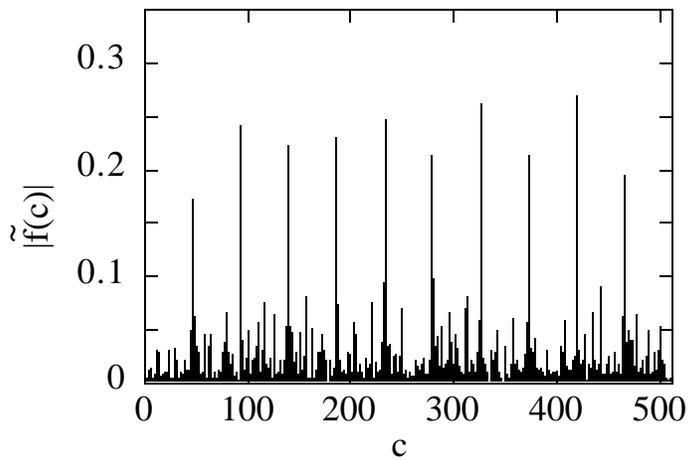
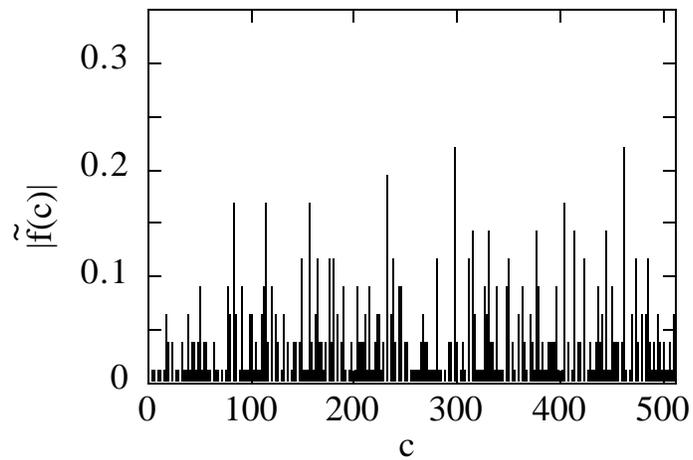
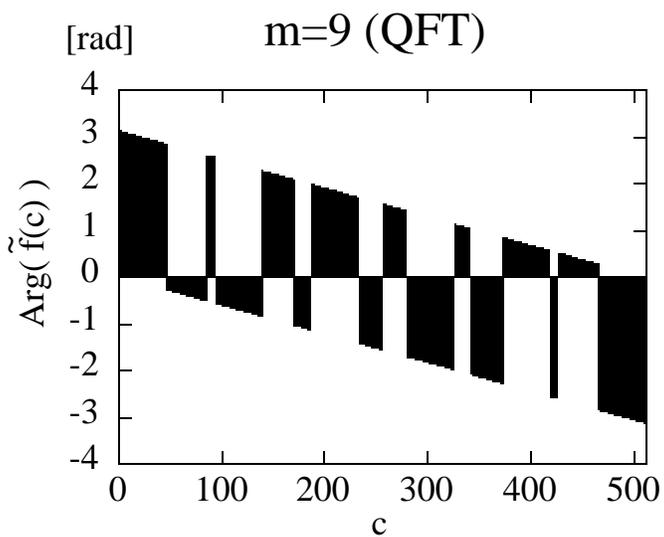
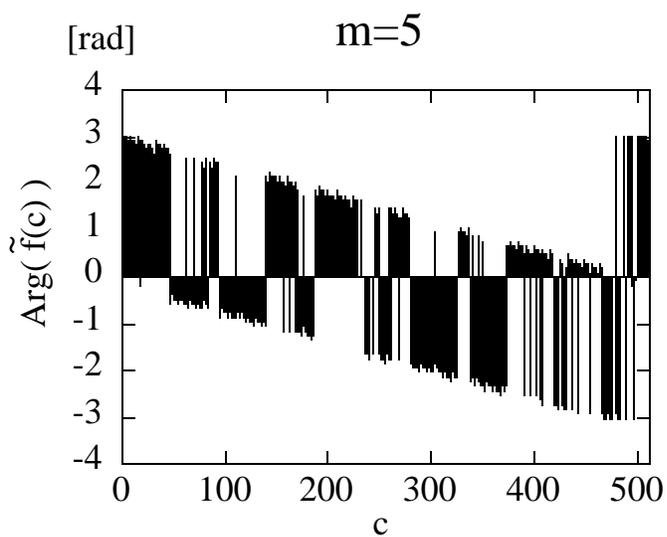
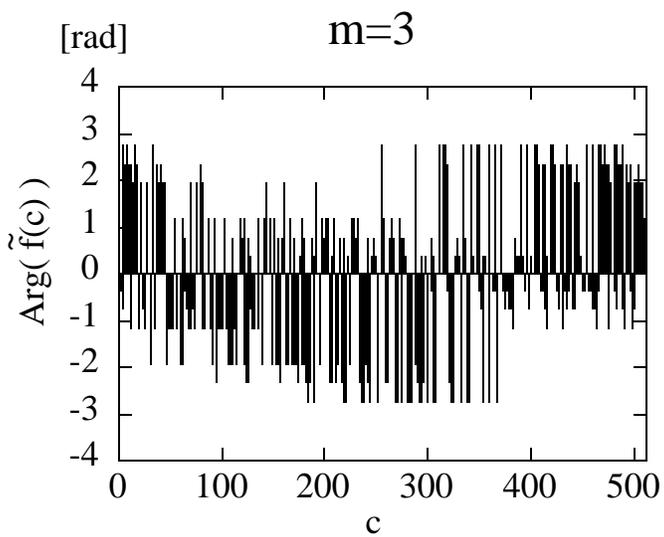
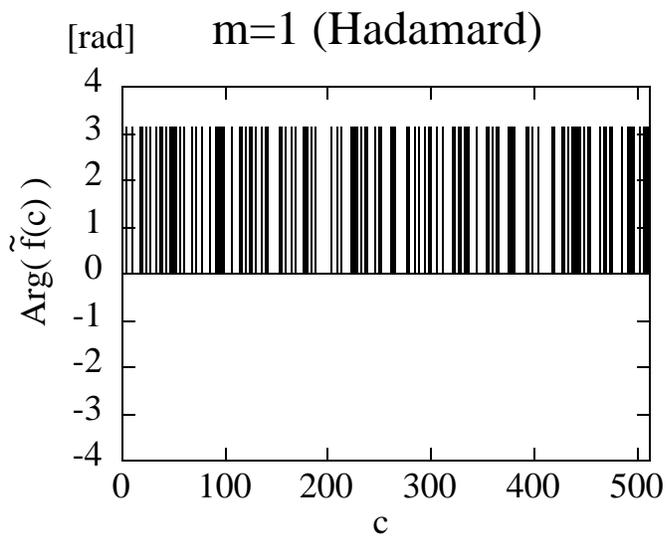
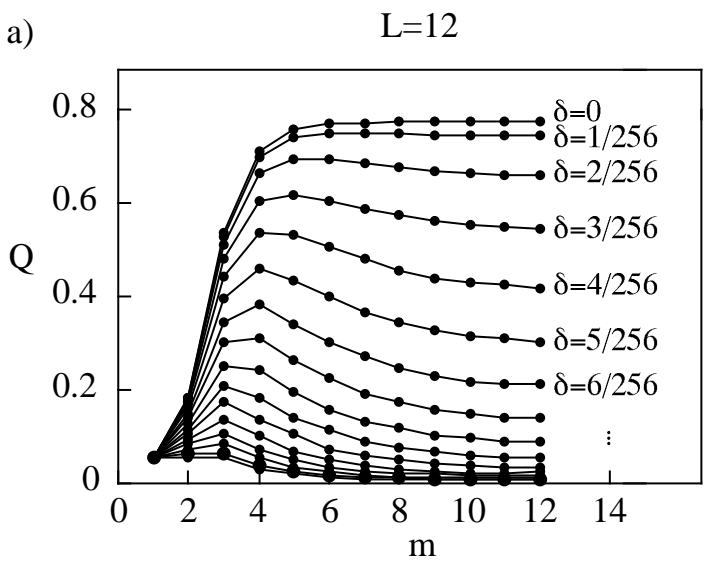
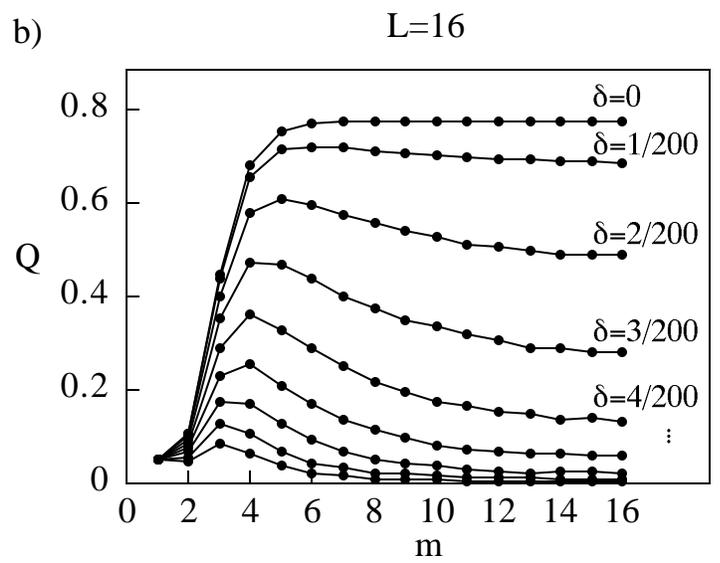
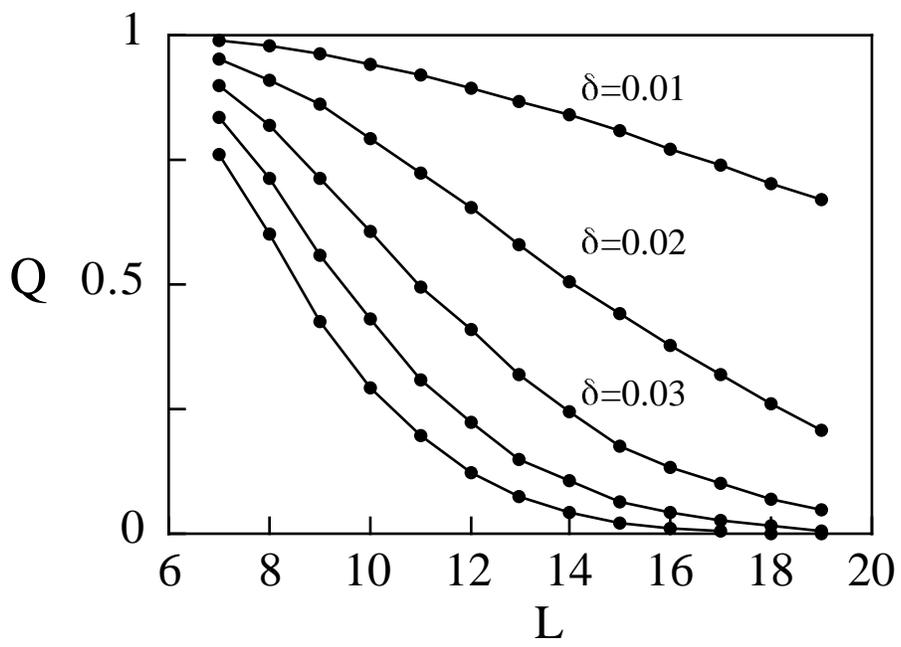
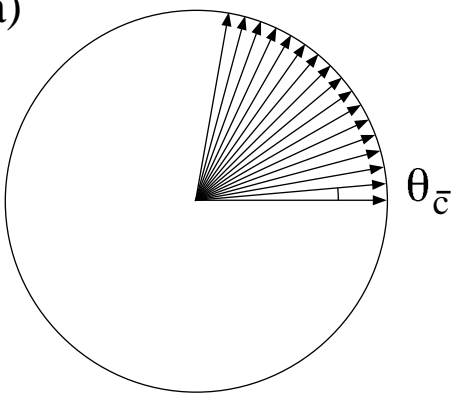
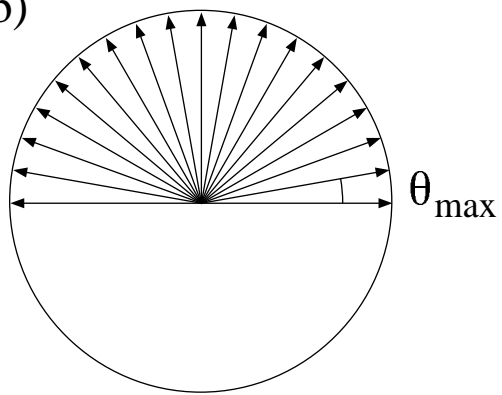
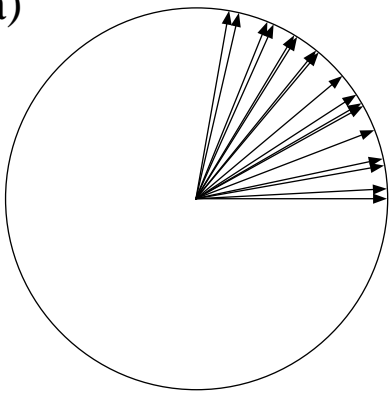
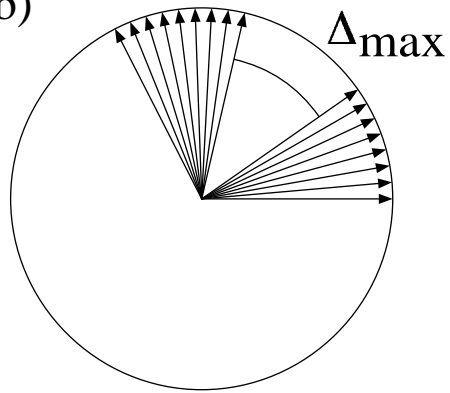
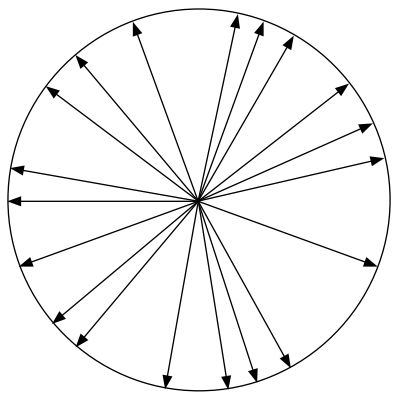
\end{document}